\documentclass[aps,prl,reprint,superscriptaddress,showpacs]{revtex4-1}
\usepackage{graphicx}
\usepackage{amsmath}
\usepackage{bm}
\usepackage{dcolumn}
\usepackage{amsfonts}
\usepackage{amssymb}
\usepackage[breaklinks=true]{hyperref}
\hypersetup{colorlinks=true, citecolor=blue, filecolor=blue, linkcolor=blue, urlcolor=blue}
\usepackage{color}

\begin{document}

\title{Criteria for accurate determination of the magnon relaxation length from the nonlocal spin Seebeck effect}
\author{J. Shan}
\email[]{j.shan@rug.nl}
\author{L. J. Cornelissen}
\author{J. Liu}
\affiliation{Physics of Nanodevices, Zernike Institute for Advanced Materials, University of Groningen, Nijenborgh 4, 9747 AG Groningen, The Netherlands}
\author{J. Ben Youssef}
\affiliation{Universit\'{e} de Bretagne Occidentale, Laboratoire de Magn\'{e}tisme de Bretagne CNRS, 6 Avenue Le Gorgeu, 29285 Brest, France}
\author{L. Liang}
\author{B. J. van Wees}
\affiliation{Physics of Nanodevices, Zernike Institute for Advanced Materials, University of Groningen, Nijenborgh 4, 9747 AG Groningen, The Netherlands}

\date{\today}

\begin{abstract}

The nonlocal transport of thermally generated magnons not only unveils the underlying mechanism of the spin Seebeck effect, but also allows for the extraction of the magnon relaxation length ($\lambda_m$) in a magnetic material, the average distance over which thermal magnons can propagate. In this study, we experimentally explore in yttrium iron garnet (YIG)/platinum systems much further ranges compared with previous investigations. We observe that the nonlocal SSE signals at long distances ($d$) clearly deviate from a typical exponential decay. Instead, they can be dominated by the nonlocal generation of magnon accumulation as a result of the temperature gradient present away from the heater, and decay geometrically as $1/d^2$. We emphasize the importance of looking only into the exponential regime (i.e., the intermediate distance regime) to extract $\lambda_m$. With this principle, we study $\lambda_m$ as a function of temperature in two YIG films which are 2.7 and 50 $\mu$m in thickness, respectively. We find $\lambda_m$ to be around 15 $\mu$m at room temperature and it increases to 40 $\mu$m at $T=$ 3.5 K. Finite element modeling results agree with experimental studies qualitatively, showing also a geometrical decay beyond the exponential regime. Based on both experimental and modeling results we put forward a general guideline for extracting $\lambda_m$ from the nonlocal spin Seebeck effect.

\end{abstract}
\maketitle

\section{I.~~~ Introduction}

Since its discovery \cite{uchida_observation_2008,uchida_spin_2010}, the spin Seebeck effect (SSE) has been a central topic in the burgeoning field of spin caloritronics \cite{bauer_spin_2010,bauer_spin_2012,boona_spin_2014}, not only due to its promising application in utilizing thermal energy on a large scale \cite{kirihara_spin-current-driven_2012}, but also because of its rich and interesting physics \cite{jaworski_spin-seebeck_2011,uchida_quantitative_2014,kikkawa_critical_2015,jin_effect_2015,vlietstra_simultaneous_2014,kehlberger_length_2015,guo_influence_2016,kikkawa_magnon_2016,cornelissen_nonlocal_2017,meier_longitudinal_2015}. When a heat current flows through magnetic insulators such as yttrium iron garnet (YIG), a pure magnonic spin current is excited without any charge currents flowing. A magnon spin accumulation is thereby built up at the boundaries of YIG \cite{duine_spintronics_2015,cornelissen_magnon_2016-1,shan_influence_2016}, which can induce a spin angular momentum flow into an adjacent platinum (Pt) layer through interfacial exchange coupling \cite{brataas_finite-element_2000,weiler_experimental_2013,xiao_transport_2015}. It can then convert into a measurable electric voltage by the inverse spin Hall effect (ISHE) \cite{saitoh_conversion_2006}. 

Due to scattering processes such as magnon-phonon interactions, the magnon spin accumulation relaxes at a rate closely related to the phenomenological Gilbert damping coefficient $\alpha$. In the diffusive magnon transport picture, the magnon relaxation length $\lambda_m$, the average distance over which magnons can propagate, can be expressed with $\alpha$ \cite{cornelissen_magnon_2016-1}. Owing to the diffusive nature of thermally excited magnons, $\lambda_m$ is thus important for the understanding of the SSE. 

So far, $\lambda_m$ has been obtained experimentally with mainly two approaches: one is the study of longitudinal SSE signals as a function of the YIG thickness $t_{\text{YIG}}$ \cite{kehlberger_length_2015,guo_influence_2016}, and the other employs a lateral nonlocal geometry, which is also referred to as the nonlocal SSE \cite{cornelissen_long-distance_2015,giles_long-range_2015,shan_influence_2016,zhou_lateral_2017,shan_nonlocal_2017,cornelissen_nonlocal_2017}. A heater and a detector are positioned on top of a YIG surface, separated by a distance $d$, and one studies how the signals decay as a function of $d$. Particularly, $\lambda_m$'s that are acquired from these two methods exhibit roughly one order of magnitude difference at room temperature, which has been ascribed to different energy spectrum of magnons probed locally and nonlocally \cite{guo_influence_2016}.

The lateral approach is experimentally more favorable in the sense that it allows the experiments to be conducted on the same YIG surface, which circumvents the possible differences among different YIG surfaces and YIG/Pt interfaces. Nevertheless, the $\lambda_m$'s reported from the lateral geometry still seem to differ by one order of magnitude in both room and lower temperatures among different groups \cite{giles_long-range_2015,cornelissen_temperature_2016-1,zhou_lateral_2017,cornelissen_nonlocal_2017}. These discrepancies should be clearly addressed despite the material quality variations.

In the lateral approach, the electrical injection of magnons through spin voltage bias \cite{cornelissen_long-distance_2015,cornelissen_magnon_2016-1} takes place only at the injector, but the thermal generation of magnons is much more nonlocal. According to the bulk SSE picture \cite{duine_spintronics_2015,cornelissen_magnon_2016-1,shan_influence_2016,rezende_magnon_2014,kehlberger_length_2015}, a thermal magnon current is excited wherever a temperature gradient ($\nabla T$) is present, which exists not only close to the heating source, but also much further away. Therefore, the decay of nonlocal SSE signals as a function of $d$ is not solely due to magnon relaxation, but also related to $\nabla T$. This behavior complicates the extraction of $\lambda_m$. Very recently, an additional decay on top of the exponential relaxation has been observed in bulk YIG films that is 500 $\mu$m in thickness, and a longer decay length scale was associated with it \cite{giles_thermally_2017}.

\begin{figure*}
	\includegraphics[width=17cm]{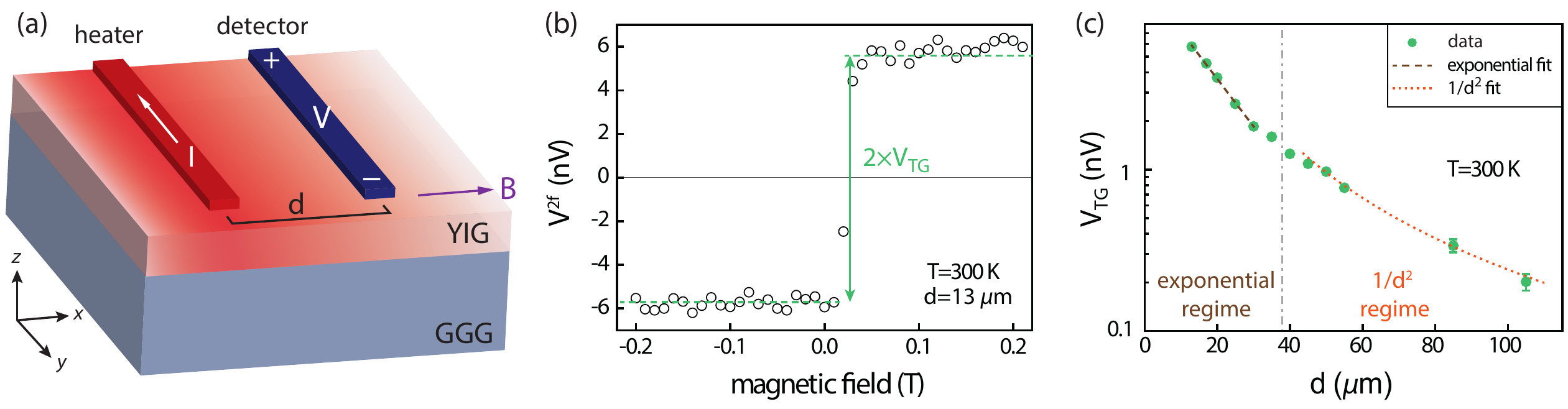}
	\caption{(a) Schematic illustration of the device structure. An ac current is sent to the heater (left Pt strip) and the voltage is detected nonlocally at the detector (right Pt strip), which is separated by a center-to-center distance $d$ from the heater. An in-plane magnetic field $B$ is applied along the $x$-axis to achieve maximal detection efficiency. (b) One typical field-sweep measurement of $V^{2f}$ performed for $d$= 13 $\mu$m at $T$= 300 K on a 2.7-$\mu$m-thick YIG film, normalized to $I=$ 100 $\mu$A, from which the amplitude of the thermally generated nonlocal signal $V_{\textup{TG}}$ can be extracted. (c) $V_{\textup{TG}}$ as a function of $d$ at $T$= 300 K for the same YIG film, plotted in a logarithmic scale. The datapoints in this plot are after the sign-reversal and are opposite in sign with the local SSE signal, and are defined as positive throughout the paper. For the datapoints in the range of 10 $\mu$m $\leq d \leq 30 \ \mu$m, they are fitted exponentially with the equation $V_{\textup{TG}}=C \exp (-d/\lambda_m)$ as shown by the brown dashed line, where $\lambda_m$= 14.7 $\pm$ 0.4 $\mu$m. In the range of 45 $\mu$m $\leq d \leq 105 \ \mu$m the datapoints are fitted with $V_{\textup{TG}}=C'/d^2$, shown by the orange dotted line. $C$ and $C'$ are coefficients that incorporate the system material properties, such as the bulk spin Seebeck coefficient $S_S$, the magnon spin conductivity $\sigma_s$ and the YIG film thickness.}	
	\label{fig1_lowT}
\end{figure*}

Despite that the electrical approach gives well-defined magnon excitation location, the nonlocal signals obtained with this approach diminish as the sample temperature is reduced \cite{cornelissen_temperature_2016-1,goennenwein_non-local_2015}, making it very difficult to study $\lambda_m$ at low temperatures. In contrast, the nonlocal signals from thermal generation often remain sufficiently large or even increase substantially at lower temperatures \cite{cornelissen_temperature_2016-1,zhou_lateral_2017}. It is hence more practical to study $\lambda_m$ with a Joule heating approach.

In this study, we investigate the nonlocal SSE signals carefully by exploring the ultra-long heater-detector distance regime, i.e., around one order larger than the typical $\lambda_m$ we found in our previous studies \cite{cornelissen_long-distance_2015,cornelissen_temperature_2016-1,shan_influence_2016}. We can then clearly distinguish two decay regimes, which are governed by two different processes: One is dominated by the relaxation of the magnon chemical potential buildup around the local heating source, where the signals exhibit an exponential decay on the length scale of $\lambda_m$; the other regime locates at a much further distance, dominated by the magnon accumulation generated nonlocally as a result of the nonzero $\nabla T$ in the vicinity of the detector, with the signals clearly deviating from an exponential decay. We found and established that they exhibited a $1/d^2$ decay manner instead. We demonstrate the complexity to study $\lambda_m$ from a thermal method, and highlight the importance to only evaluate the proper regime to obtain $\lambda_m$.

Furthermore, we carry out a systematic study at a wide range of temperatures, and find that the magnon exponential regime extends to a further distance as $\lambda_m$ becomes larger at lower temperatures ($T <$ 20 K). By exponential fitting only the magnon exponential regime we reliably extract $\lambda_m$ ranging from 3.5 K to 300 K. Finally, we perform finite element modeling with various $\lambda_m$, which yields consistent results that support our understanding by showing also different decay regimes, with the same decay manners as observed experimentally. We conclude with a general rule for extracting $\lambda_m$ in nonlocal SSE studies.

\section{II.~~~ Experimental details}

In the present study, we use YIG (111) films with two different thicknesses, 2.7 $\mu$m and 50 $\mu$m, both grown by liquid phase epitaxy on single-crystal Gd$_{3}$Ga$_{5}$O$_{12}$(GGG) (111) substrates.
The 50-$\mu$m-thick YIG sample was purchased from Matesy GmbH, and the 2.7-$\mu$m-thick YIG sample was provided by the Universit\'{e} de Bretagne in Brest, France. Pt strips (6.5 $\pm$ 0.5 nm in thickness, 100 $\mu$m and 1 $\mu$m in length and width, respectively) aligned in parallel directions with distance $d$ relative to each other were patterned by electron beam lithography and sputtered onto a YIG substrate, as schematically shown in Fig.~\ref{fig1_lowT}(a). Multiple devices were fabricated with various $d$ on a single substrate. Contacts consisting of Ti (5 nm)/Au (75 nm) were subsequently patterned and evaporated to connect the Pt strips. 

Compared to our previous experiments on this YIG substrate \cite{shan_influence_2016}, the Pt strips were designed to be wider and longer in this study for two main reasons. First, with wider strips one can send larger currents through, which significantly improves the signal-to-noise ratio, making it possible to probe the small signals in the long-$d$ regime. Second, longer strips reduce the effects of magnon currents that leak away in the $y$-axis direction, allowing for a 2D analysis in the $x$-$z$ plane.

The samples were measured by sweeping the magnetic field along the $x$-axis. A lock-in detection technique is used, where an ac current $I$, typically with a frequency of 13 Hz and an rms value of 100 $\mu$A, was sent through one of the Pt strips (the heater), and the voltage output was monitored nonlocally at the other Pt strip (the detector). In this study, we focus on the behavior of the thermally excited magnons, which results from Joule heating at the heater and is hence a second-order effect with respect to $I$. This is captured in the second harmonic signals $V^{2f}$ in the lock-in measurement, as $V^{2f}=\frac{1}{\sqrt{2}}I_0^2 \cdot R_2$ with a phase shift of -90$^{\circ}$ provided no higher even harmonic signals are present.  The data plotted in this paper were all normalized to $I$=100 $\mu$A. The samples were placed in a superconducting magnet cryostat with a variable temperature insert to enable temperature-dependent measurements, ranging from 3.5 K to 300 K in this study. The sample temperature is always checked to be fully stabilized before performing measurements on all devices at that specific temperature. Furthermore, the applied charge current $I$ is ensured to be in the linear regime, such that the Joule heating does not increase the average device temperature significantly.

\section{II.~~~Results and Discussion}

\subsection{A. ~~~Results on 2.7-$\mu$m-thick YIG}

\subsubsection{1. ~~~Room temperature results}

 A typical field-sweep measurement curve is shown in Fig.~\ref{fig1_lowT}(b). From the ISHE, one gets a maximum signal when the YIG magnetization is perpendicular to the Pt detector strip. Reversing the YIG magnetization results in an opposite polarization of the magnon spin current and consequently a reverse sign of the signal. As the employed YIG films have very small coercive fields \cite{vlietstra_spin-hall_2013}, the signal jump around zero field allows us to extract the amplitude of the thermally generated nonlocal signal $V_{\text{TG}}$. We focus on the low-field regime where the magnetic-field-induced SSE suppression  \cite{kikkawa_critical_2015,cornelissen_magnetic_2016} can be excluded in our analysis.
 
 To study how the signals decay laterally, we further measured $V_{\text{TG}}$ for all devices and plot them as a function of $d$, as shown in Fig.~\ref{fig1_lowT}(c).  Note that the shortest distance we probed here ($d=$ 10 $\mu$m) is already further than the sign-reversal distance $d_{\text{rev}}$ for the 2.7-$\mu$m-thick-YIG, around 5 $\mu$m at room temperature \cite{shan_influence_2016}, so that the sign of $V_{\text{TG}}$ in this study is opposite to the sign of the local spin Seebeck signal, which is obtained with the heater itself as the detector. In the beginning, the signals follow an exponential decay, where $\lambda_m$= 14.7 $\pm$ 0.4 $\mu$m can be extracted. This is the ``relaxation regime'' described in Ref.~\cite{cornelissen_long-distance_2015}. Here we name it as ``exponential regime''. The signals at further distances, however, clearly deviate from this exponential fit. They exhibit a slower decay, which can be well fitted with a $1/d^2$ function. Here we name it as ``$1/d^2$ regime''.
 
\begin{figure}[t]
 	\includegraphics[width=8.5cm]{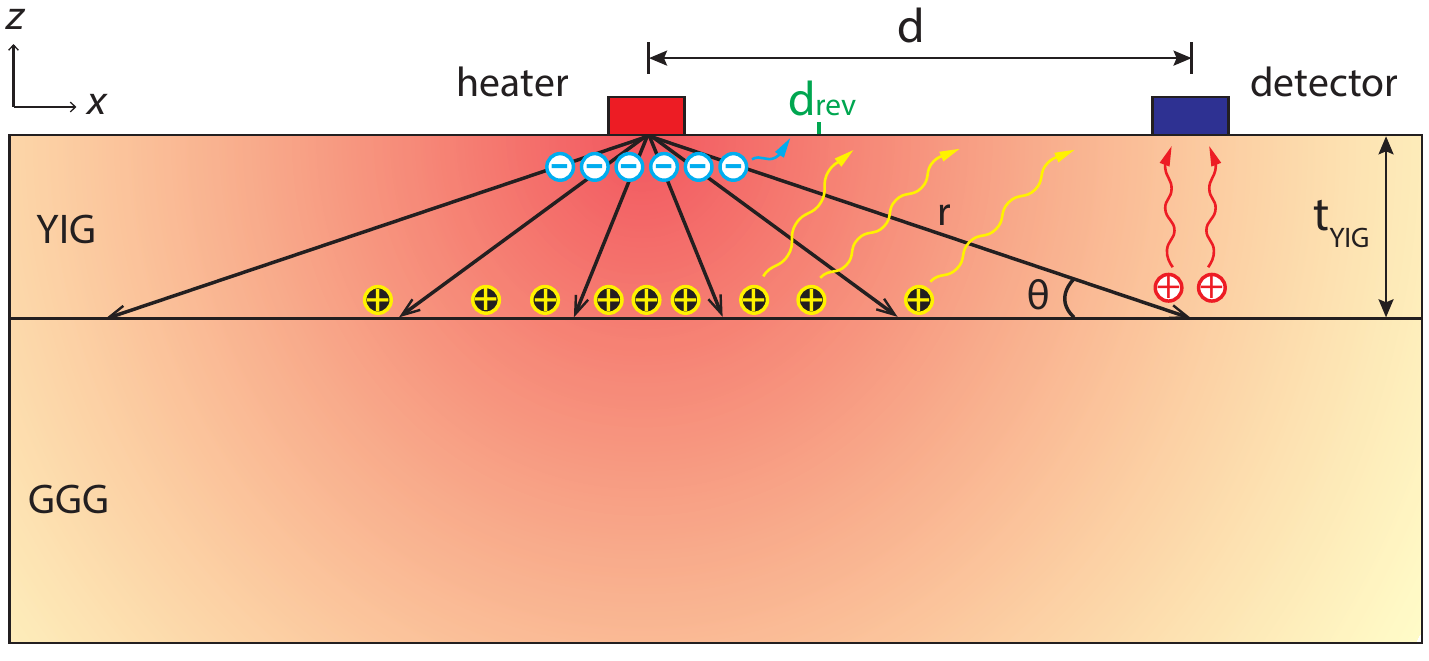}
 	\caption{Schematic cross-section view of the device in the $xz$-plane. A charge current flows through the heater and generates a radial temperature gradient profile in both YIG and GGG layers, centered around the heater, as illustrated with the background color. A thermal magnon flow (represented by black arrows) is induced along the same direction as the heat flow in the YIG layer, as a result of the SSE. Unlike the heat flow, the magnon flow cannot enter the GGG layer, and a magnon accumulation (indicated by the `+' sign) is therefore built up at the YIG/GGG interface. Likewise, at the YIG/heater interface, a magnon depletion (indicated by the `-' sign) is formed. For the magnon accumulation, the yellow circles indicate the generation below the heater, and the red circles indicate the nonlocal generation near the detector.}	
 	\label{fig2_lowT}
 \end{figure}

 According to our previously proposed SSE picture \cite{cornelissen_magnon_2016-1,shan_influence_2016,cornelissen_nonlocal_2017}, the heat flow $J_q$ sourced from the heater induces a thermal magnon flow $J_{m,q}$ along with it inside the YIG layer. When $J_{m,q}$ reaches the YIG/GGG interface, it cannot enter further into the GGG layer. Because of this abrupt change in magnon spin conductivity, a magnon accumulation (corresponding to a positive magnon chemical potential, $\mu_m^+$) is formed at the bottom of the YIG layer, as shown in Fig.~\ref{fig2_lowT}. Similarly, a magnon depletion (corresponding to a negative magnon chemical potential, $\mu_m^-$) is formed at around the heater. As a consequence, the gradient of $\mu_m$ drives a diffusive magnon flow $J_{m, \text{diff}}$ to counteract $J_{m,q}$, such that the boundary conditions are satisfied (in this case an open-circuit condition for spin currents at the bottom interface of YIG, and at the top of YIG the boundary condition depends on the spin opacity of the YIG/heater interface \cite{shan_influence_2016}).
 
 \begin{figure*}
	\includegraphics[width=17cm]{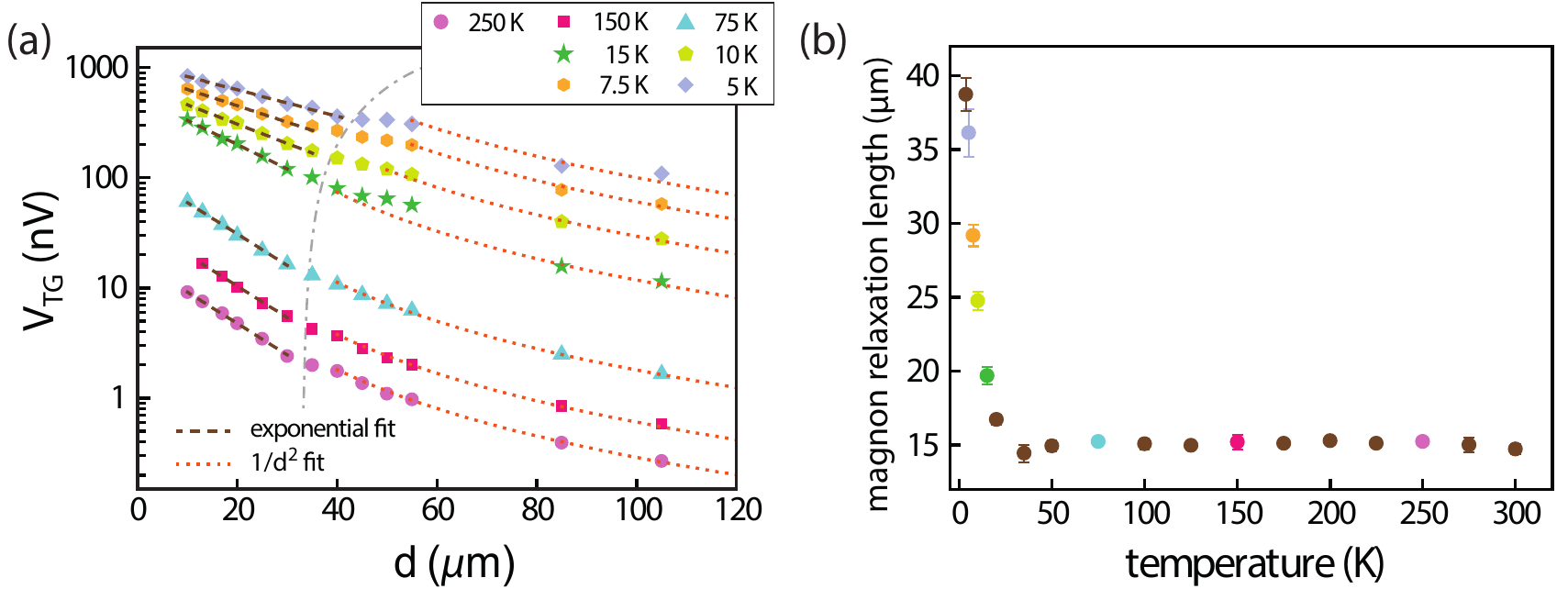}
	\caption{(a) Measured $V_{\text{TG}}$ as a function of $d$ for various temperatures on a 2.7-$\mu$m-thick YIG film. The exponential and quadratic decay fits are performed in a similar fashion as in Fig.~\ref{fig1_lowT}. For $T <$ 20 K, due to the increased $\lambda_m$, the exponential regimes extend to longer $d$, and consequently quadratic decay regimes start at further distances. But for the sake of consistency, the $\lambda_m$'s are all determined from exponential fits  performed on the datapoints within 10 $\mu$m $\leq d \leq 30 \ \mu$m. (b) $\lambda_m$'s extracted from exponential fits at temperatures from $T=3.5$ K to 300 K.}	
	\label{fig3_lowT}
\end{figure*} 

Because of the radial shape of the temperature profile, $\mu_m^-$ is present close to the heater, surrounded by $\mu_m^+$ that extends further away. The relative position of the two, or essentially the zero-crossing line of $\mu_m$, is influenced by $t_{\text{YIG}}$ and heater spin opacity among others \cite{shan_influence_2016}.  After the sign reversal, $\mu_m^+$ first grows to its maximum, and then diffuses in the lateral direction, relaxing exponentially on the length scale of $\lambda_m$. This can be mapped by the ISHE signal produced by the Pt detector, which reflects the $\mu_m$ along the YIG surface. $\lambda_m$ can be extracted by fitting the obtained signals in the exponential regime by an exponential decay \cite{shan_influence_2016,cornelissen_long-distance_2015,cornelissen_nonlocal_2017}.

The determination of $\lambda_m$ from data before the sign-reversal \cite{giles_long-range_2015,shan_influence_2016,giles_thermally_2017}, i.e., checking the relaxation of the $\mu_m^-$, is also possible, but only valid when $t_{\text{YIG}} \gg \lambda_m$. This issue will be further discussed in subsections B - D.

It should be noted, however, that at very long distances where $\mu_m^+$ diffusing from around the heater becomes almost zero due to magnon spin relaxation, there can still be a small $\nabla T$ present at the YIG/GGG interface below the detector. Within the same framework of the bulk SSE picture, this will induce a thermal magnon flow $J_{m,q}$ proportional to it, building $\mu_m^+$ due to the open-circuit condition. A $J_{m,\text{diff}}$ driven by it can therefore diffuse into the detector and convert into a signal, as shown in Fig.~\ref{fig2_lowT}. Note that we do not assume the Pt detector to be a heat sink so that there is no heat current flowing into the Pt detector, but the detected magnon current is diffused from the YIG/GGG interface beneath it.  

The signals at long distances hence decay independent of $\lambda_m$. To derive how they decay as a function of $d$, for simplicity we first assume that the thermal conductivities of YIG and GGG, $\kappa_{\text{YIG}}$ and $\kappa_{\text{GGG}}$, are similar in value such that the heat flows radially even when $d > t_{\text{YIG}}$. At a certain $d$, the magnitude of the $J_q$ that crosses the YIG/GGG interface is then proportional to $1/\pi r$, with $r=\sqrt{d^2+t_{\text{YIG}}^2}$.
$J_{m,q}$ reaches the bottom of the YIG layer at an angle $\theta$, where $ \theta = \arctan(t_{\text{YIG}} /d)$, as shown in Fig.~\ref{fig2_lowT}.  Yet only the part of $J_{m,q}$ that is normal to the YIG/GGG interface would encounter the GGG barrier and generates a $\mu_m^+$:
\begin{equation}
J_{m,q}^{z} \propto \frac{1}{\pi r} \cdot \sin \theta
%=\frac{t_{\textup{YIG}}}{\pi r^2}
=\frac{t_{\textup{YIG}}}{\pi(d^2+t_{\textup{YIG}}^2)} 
\stackrel{t_{\textup{YIG}} \ll d} 
{\approx} \frac{t_{\textup{YIG}}}{\pi d^2}
\label{eq:Jmq_normal} .
\end{equation}
The resulting $\mu_m^+$ would then induce a diffusive magnon flow proportional to $J_{m,q}^{z}$, which can enter the detector at $d$. This explains the $1/d^2$ dependence of $V_{\text{TG}}$. Note that the signal at the detector $V_{\text{TG}}$ is not necessarily proportional to $t_{\text{YIG}}$, as the relaxation from the bottom to the top side of YIG needs to be taken into account, unless $t_{\text{YIG}}$ is much smaller than $\lambda_m$. 

For the relation in Eq.~\ref{eq:Jmq_normal} to hold, $\kappa_{\text{YIG}}$ does not have to be strictly equal to $\kappa_{\text{GGG}}$. The $1/d^2$ dependence is in general valid as long as $\kappa_{\text{YIG}} \leq \kappa_{\text{GGG}}$. In this case, the heat flow towards the GGG layer dominates the one that remains in the YIG layer, and an increase of $d$ would result in a decrease of $J_{m,q}$ in a nearly $1/d$ manner and hence $J_{m,q}^z$ a $1/d^2$ manner. In fact, the smaller the ratio of $\kappa_{\text{YIG}}$ over $\kappa_{\text{GGG}}$, the more accurate the approximation in Eq.~\ref{eq:Jmq_normal} is. Conversely, when $\kappa_{\text{YIG}} \gg \kappa_{\text{GGG}}$, the relation in Eq.~\ref{eq:Jmq_normal} is no longer valid. 

\subsubsection{2. ~~~Results at low temperatures}

We further performed the same measurements at various temperatures on 2.7-$\mu$m-thick YIG, in order to study $\lambda_m$ carefully as a function of temperature, as well as to confirm the above picture. 

The main results are shown in Fig.~\ref{fig3_lowT}. As shown in Fig.~\ref{fig3_lowT}(a), the $V_{\text{TG}}$ for all distances enhance when decreasing the temperature, consistent with the general trend in our previous results on 0.21-$\mu$m-thick YIG film \cite{cornelissen_temperature_2016-1}. However, in this study we do not observe reductions of $V_{\text{TG}}$ below 7 K as in Ref.~\cite{cornelissen_temperature_2016-1}, which could be due to the subtle differences between the employed YIG films in both studies and still requires further investigation.

For almost all temperatures at which measurements are carried out, $V_{\text{TG}}$ apparently cannot be fitted by a single exponential decay, similar to the observation at room temperature. Following the same procedure, we separate the data into two regimes and fit them into exponential and quadratic decay, respectively. 

The extracted $\lambda_m$'s from the exponential fits across the whole temperature range are shown in Fig.~\ref{fig3_lowT}(b). One can see that down to $T$= 35 K, $\lambda_m$ remains more or less unchanged as a function of temperature. This is also in line with our previous study on 0.21-$\mu$m-thick YIG film \cite{cornelissen_temperature_2016-1}. At $T <$ 20 K, however, we observe a sharp and monotonic increase of $\lambda_m$ when reducing temperature. Consequently, the transition between the two decay regimes extend to a longer $d$, as the diffused magnon accumulation can be further preserved.

The $1/d^2$ decay can be fitted satisfactorily at long distances even down to very low temperatures. From literature, both $\kappa_{\text{YIG}}$ and $\kappa_{\text{GGG}}$ of bulk materials vary by more than one order of magnitude from room temperature to their peak values, which take place roughly between 20 K and 30 K \cite{slack_thermal_1971,hakuraku_thermal_1983,iguchi_concomitant_2017}. Yet the general shapes of $\kappa_{\text{YIG}}$ and $\kappa_{\text{GGG}}$ as a function of temperature are very similar. Additionally, for YIG thin films, the thermal conductivities are found to be smaller than their bulk values \cite{euler_thermal_2015}. Therefore, we can say that in the measured temperature range, $\kappa_{\text{YIG}} \leq \kappa_{\text{GGG}}$ should hold according to literature values. 

\subsubsection{3. ~~~2D Comsol modeling results}

We perform next numerical modeling that solves profiles of the temperature and $\mu_m$ in our studied system using a Comsol model. From the model we can calculate $V_{\text{TG}}$ for even further $d$ than studied experimentally, which allows us to identify and study the different decay regimes more clearly.

We use a two-dimensional finite element model as already described in detail in Ref.~\cite{shan_influence_2016}. Except for a few geometrical parameters, such as Pt strip widths, Pt and YIG film thicknesses, the physics and the rest of the material parameters are kept to be the same as in Ref.~\cite{shan_influence_2016} for the sake of consistency. The focus of the numerical study in this section, however, is the modeled signals in the $1/d^2$ regime, which has not been investigated so far. 

We do not aim for quantitative agreement between the experimental and modeled results, as in the model we only vary the input $\lambda_m$, while in reality, the change of temperature does not only evoke the variation of $\lambda_m$, but also other crucial parameters such as $\kappa_{\text{YIG}}$ and $\kappa_{\text{GGG}}$, the magnon spin conductivity of YIG \cite{cornelissen_temperature_2016-1}, the effective spin mixing conductance at the YIG/Pt interface and the spin Seebeck coefficient of YIG \cite{cornelissen_magnon_2016-1}, etc. The absolute magnitudes of $V_{\text{TG}}$ and the exact starting and ending distances of the exponential regimes, cannot be directly compared between the experimental and modeled results without several assumptions. Nevertheless, the model works qualitatively, so that the decay manner of $V_{\text{TG}}$ can be studied and compared with experimental results.

\begin{figure}[t]
	\includegraphics[width=8.5cm]{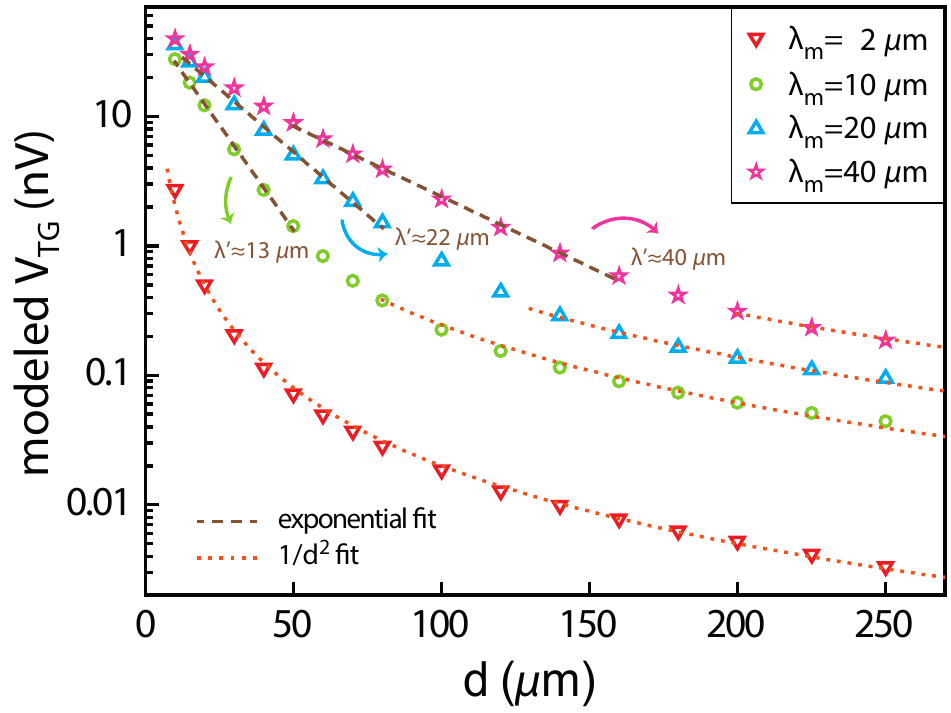}
	\caption{Modeling results of the nonlocal SSE signals on a 2.7-$\mu$m-thick YIG film in the range of 10 $\mu$m $\leq d \leq  250\  \mu$m, with different $\lambda_m$ as modeling input, while all the other parameters are kept unvaried. The extracted length scales $\lambda$' by exponential fittings are indicated nearby. All the modeled signals presented here are after the sign-reversal distance.}	
	\label{fig4_lowT}
\end{figure}

Fig.~\ref{fig4_lowT} shows the modeled $V_{\text{TG}}$ as a function of distance up to $d=250 \  \mu$m. We calculated the signals for different magnon relaxation length input $\lambda_m$ to check the dependence of the two decay regimes on $\lambda_m$. The datapoints at very short distances before the sign reversal are not plotted here, as they are not of central interest in this study. 

\begin{figure*}[t]
	\includegraphics[width=16cm]{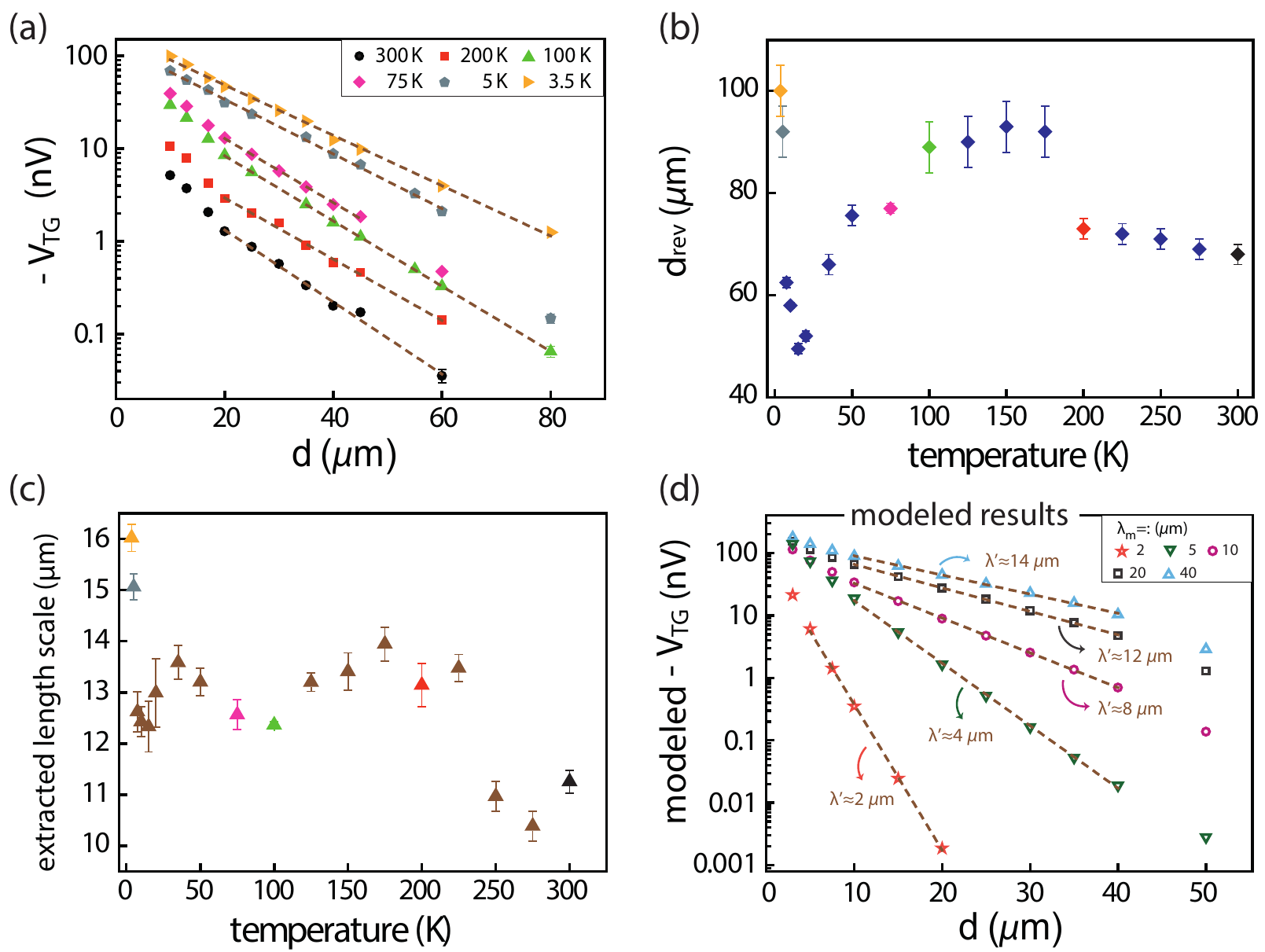}
	\caption{Experimental and modeling results on a 50-$\mu$m-thick YIG film.  (a) $-V_{\text{TG}}$ as a function of $d$ for various temperatures. Note that the sign of all datapoints plotted here are the same as the local SSE signal, which we define as negative. Only the datapoints before the sign-reversal are shown in this plot. Brown dashed lines are exponential fittings similar as described in Fig.~\ref{fig1_lowT}, with the pre-exponential coefficients $C$ being opposite in sign. (b) The sign-reversal distances obtained by interpolation ($d_{\text{rev}}<$ 80 $\mu$m) and extrapolation ($d_{\text{rev}}>$ 80 $\mu$m) for different temperatures. (c) The extracted length scales (not necessarily equal to $\lambda_m$) from exponential fits from $T=$ 3.5 K to 300 K. (d) The modeled $V_{\text{TG}}$ for different input $\lambda_m$, with extracted length scales $\lambda$' indicated nearby.}	
	\label{fig5_lowT}
\end{figure*}

The modeled results reproduce the shapes of the experimental data quite well. The signals first exhibit an exponential decay, where the starting and ending distances depend on $\lambda_m$, and then followed by a $1/d^2$ decay. For $\lambda_m=2 \ \mu$m, the exponential regime is too short and takes place before $d=10 \  \mu$m, and therefore not captured in this plot. Instead, $1/d^2$ decay dominates the full investigated distance range. 

One can also obtain the extracted magnon relaxation length $\lambda$' by fitting the exponential regimes. $\lambda$' is very close to the input $\lambda_m$, which justifies the way we extracted $\lambda_m$ in Fig.~\ref{fig3_lowT}. 

\subsection{B. ~~~Results on 50-$\mu$m-thick YIG}

We now show a set of measurements on a 50-$\mu$m-thick YIG film. Similar devices as on 2.7-$\mu$m-thick YIG film were fabricated with $d$ ranging from 10 $\mu$m to 80 $\mu$m.

In Ref.~\cite{shan_influence_2016} we have already investigated $d_{\text{rev}}$ of this YIG film at room temperature, which takes place between $d=$ 60 $\mu$m and $d=$ 80 $\mu$m. In this study, we look at how the nonlocal SSE signals evolve at lower temperatures.

Fig.~\ref{fig5_lowT}(a) shows the $V_{\text{TG}}$ as a function of $d$ before the sign-reversal for various temperatures on a logarithmic scale. Except for the datapoints that are still close to the heater or close to the sign-reversals, the rest of the datapoints decay exponentially. The $d_{\text{rev}}$ for each measured temperature is obtained by either interpolation or extrapolation, as shown in Fig.~\ref{fig5_lowT}(b). The general trend of $d_{\text{rev}}$ is similar as reported in Ref.~\cite{ganzhorn_temperature_2017} down to $T$=15 K, where much thinner YIG films were investigated. However, we observed a clear upturn below $T$=15 K, which seems to correspond to the upturn of the increased $\lambda_m$ as discussed below.

The length scales that are extracted from exponential fittings are shown in Fig.~\ref{fig5_lowT}(c). However, the length scales extracted before the sign-reversal can underestimate the real $\lambda_m$ if $d_{\text{rev}}$ falls in the exponential regime, which can happen when $t_{\text{YIG}}$ is comparable to $\lambda_m$. This can be true for low temperatures where $\lambda_m$ greatly increases.

To see how much we could possibly undervalue $\lambda_m$, we perform finite element modeling similar as above in Fig.~\ref{fig4_lowT}, and check the results for different $\lambda_m$. For the modeling here, we adjusted two parameters to better fit the sign-reversal. The YIG spin conductivity was increased from $5 \times 10^5$ S/m to $5 \times 10^6$ S/m and the YIG/Pt interface conductivity was decreased from $9.6 \times 10^{12}$ S/m$^2$ to $1 \times 10^{12}$ S/m$^2$. This modification does not influence the qualitative behavior of the nonlocal SSE signals.

\begin{figure}[t]
	\includegraphics[width=8.5cm]{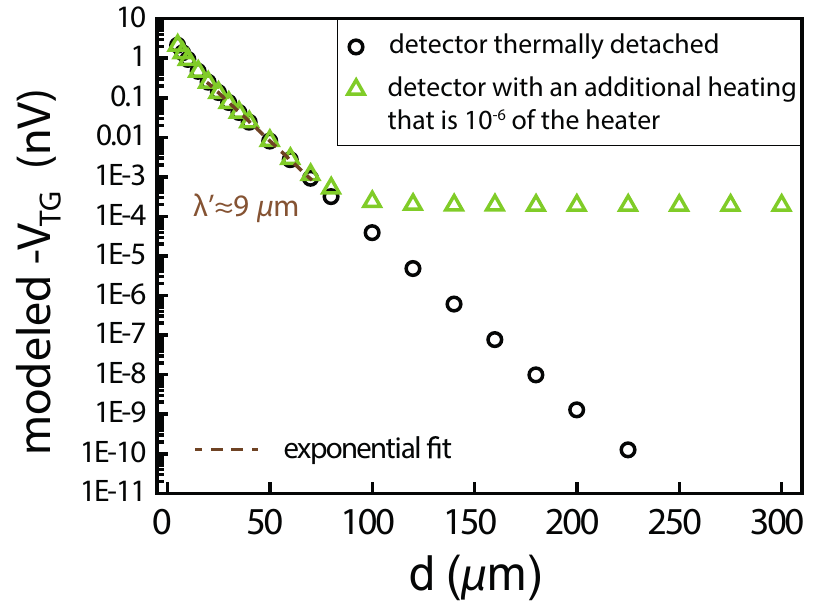}
	\caption{Modeling results of the nonlocal SSE signals on a bulk YIG material (450 $\mu$m in thickness) in the range of 5 $\mu$m $\leq d \leq  300\  \mu$m. Black circles show a single exponential decay, with the detector thermally uncoupled from YIG. Green triangles show the situation when additional Joule heating (one millionth of the amount of the heating power in the heater) is added to the detector, deviating the signals significantly in the long-distance regime. All the modeled signals presented here are before the sign reversal.}	
	\label{fig6_lowT}
\end{figure}

We fit the modeled $V_{\text{TG}}$ exponentially and obtain the corresponding length scales $\lambda$', as indicated in the figure. One can see that for $\lambda_m$=2 $\mu$m, we could extract a $\lambda$' which equals to $\lambda_m$. As $\lambda_m$ is longer, the condition $t_{\text{YIG}} \gg \lambda_m$ gradually becomes invalid, and the deviation of $\lambda$' from $\lambda_m$ gets larger.

It is therefore reasonable to assume that the extracted length scales in Fig.~\ref{fig5_lowT}(b) are only valid at higher temperatures, while at lower temperatures the real $\lambda_m$'s can be longer than extracted ones. Considering the model shows more than a factor of 2 between $\lambda_m$ and $\lambda$' when $\lambda_m =$ 40 $\mu$m, it is highly possible that, for instance, the real $\lambda_m$ reaches around 30 to 40 $\mu$m at $T=$ 3.5 K, which is consistent with the results obtained from the 2.7-$\mu$m-thick YIG film as shown in Fig.~\ref{fig3_lowT}(b). However, experimentally it is very difficult to obtain the real $\lambda_m$ for this thickness with the SSE method at very low temperatures.

\subsection{C.~~~Modeling results on bulk YIG}

For the sake of completeness, we further model the nonlocal SSE signals for a bulk YIG sample, as employed in a recent experiment \cite{giles_thermally_2017}. For such a thick YIG material, the sign reversal takes place much further than the normal studied distances, and the extraction of $\lambda_m$ becomes again possible in the exponential regime. We do not expect the $1/d^2$ decay to play a significant role, as it should only show up after the sign-reversal. Yet it was shown both in the model and experiment that a deviation from the exponential decay can be observed at longer distances, caused by the presence of a $\nabla T$ close to the detector \cite{giles_thermally_2017}.

In the simulation, when we thermally detach the detector by setting the thermal conductivity of the detector/YIG interface to zero, the modeling results show a single exponential decay based on $\lambda_m$, as shown by the black circles. This suggests that the deviation is indeed caused by the unwanted heat current flowing into or out of the detector. To show to which extent the detector signals can be influenced, we intentionally introduce a Joule heating into the detector which amounts to $10^{-6}$ of the power in the injection heater, with the detector thermally coupled with YIG. The results are shown by the green triangles in Fig.~\ref{fig6_lowT}, indicating that even very small heat flows would strong affect the signals at long distances.

These results show that in bulk YIG materials, one should extract $\lambda_m$ by only investigating the exponential regime, whereas the datapoints beyond this regime should also be excluded. However, another length scale is not necessary to be included to describe the long-$d$ behavior of the signals.

\subsection{D.~~~Summary}

\begin{figure}[t]
	\includegraphics[width=8.5cm]{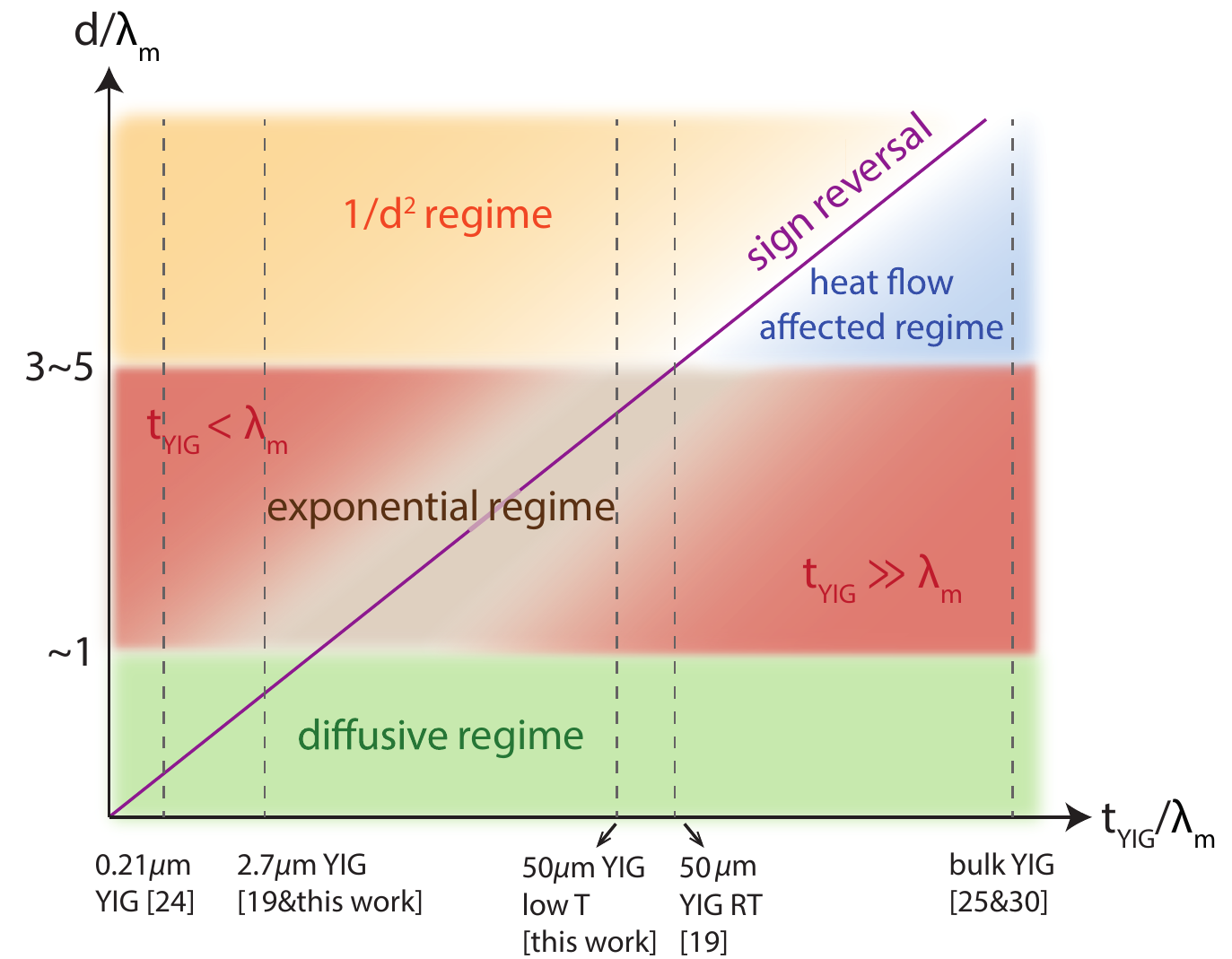}
	\caption{Schematic diagram showing different regimes for nonlocal SSE signals and the general rule for extracting $\lambda_m$ using the thermal method. The purple line indicates the sign reversal, with the location $d_{\text{rev}}$ linearly depending on $t_{\text{YIG}}$. Determination of $\lambda_m$ should be performed only in the exponential regime and far away from the sign reversal, as indicated by the red-shaded areas. Blue-shaded area denotes the deviation from exponential regime caused by heat flowing into the detector.}	
	\label{fig7_lowT}
\end{figure}

Based on the results from both YIG samples as well as previous results \cite{cornelissen_long-distance_2015,shan_influence_2016} and modeling results, we map out a general diagram for different regimes in nonlocal SSE signals, as shown in Fig.~\ref{fig6_lowT}. We consider three lengths, with $d$ and $t_{\text{YIG}}$ being geometrical lengths and $\lambda_m$ being the system parameter. 

In very short distances ($d<\lambda_m$), the system is in the diffusive regime, where the signals drop typically faster than the exponential decay \cite{cornelissen_long-distance_2015,shan_influence_2016}. In the subsequent intermediate distances, the signals decay exponentially if the sign reversal is outside this regime. If there is no overlap between the relaxations of $\mu_m^+$ and $\mu_m^-$, then one can extract $\lambda_m$ accurately from the decay of one of them, as indicated by the red zones in Fig.~\ref{fig6_lowT}. Lastly, in very long distances ($d \gg \lambda_m$) the system enters the $1/d^2$ regime, where the signal reduction no longer depends on $\lambda_m$. But for bulk YIG materials, the long-distance range deviates from the exponential regime because of the heat flow into the detector, which is distinct from the $1/d^2$ regime.

One should hence be very careful in extracting $\lambda_m$ from the lateral decay of the nonlocal SSE signal. Here we put forward a general rule of thumb to determine $\lambda_m$: One should only fit the datapoints in the exponential regime. $t_{\text{YIG}}$ should be chosen such that the sign reversal takes place outside the exponential regime. Hence, $t_{\text{YIG}}$ should be either very thin, such that the $d_{\text{rev}}< \lambda_m$ with the exponential decay reflects the relaxation of $\mu_m^+$ \cite{cornelissen_long-distance_2015}, or it should be so thick that $d_{\text{rev}} \gg \lambda_m$, and the exponential decay reflects the relaxation of $\mu_m^-$.  \cite{giles_thermally_2017,giles_long-range_2015}.

If the datapoints from the ultra-far distances are mistakenly evaluated and fitted to an exponential decay, the fitting procedure will result in an overestimation of $\lambda_m$. For YIG films where the $1/d^2$ decay dominates the ultra-far distances, the overestimated $\lambda_m$ will converge to $d_{\text{long}}/2$, where $d_{\text{long}}$ is the longest distance included in the fit. It is therefore crucial to look only at the proper regime when determining $\lambda_m$.

\section{IV.~~~Conclusions}

We studied the nonlocal SSE signals in a wide distance and temperature range. We find that for thin YIG films such as 2.7 $\mu$m in thickness, the signals exhibit first an exponential decay after the sign reversal, from which the magnon relaxation length can be estimated. Then they show a $1/d^2$ decay, due to the nonlocal generation of magnon accumulation by temperature gradient at the YIG/GGG interface near the detector. This observation further confirms the bulk generation mechanism of the SSE, and highlights the ultra-far distance detection of the nonlocal SSE signals assisted by thermal transport. We emphasize the delicate procedure to accurately obtain the magnon relaxation length from the thermally generated nonlocal signals, i.e., only the exponential regime should be investigated, with the sign reversal being far from it.

Combining our previous results on 0.21-$\mu$m-thick YIG films \cite{cornelissen_long-distance_2015,cornelissen_temperature_2016-1} and the study of this paper, we found that at room temperature $\lambda_m$'s are comparable between 0.21-$\mu$m-thick and 2.7-$\mu$m-thick YIG films, being around 9 $\mu$m and 15 $\mu$m, respectively, and in both cases they almost do not vary as a function of $T$ above 20 K. However, at very low temperatures ($T<20$ K), the $\lambda_m$ extracted from the 0.21-$\mu$m-thick YIG film does not exhibit a sharp upturn as the 2.7-$\mu$m-thick YIG film, which grows to 40 $\mu$m at  $T=$ 3.5 K. Explanation for this different behavior on these two samples requires further investigation.

\section{acknowledgments}

We thank prof.~Gerrit Bauer and dr.~Timo Kuschel for helpful discussions, M. de Roosz, H. Adema, T. Schouten and J.G. Holstein for technical assistance. This work is part of the research program of the Foundation for Fundamental Research on Matter (FOM) and is supported by NanoLab NL, EU FP7 ICT Grant InSpin 612759, NanoNextNL and the Zernike Institute for Advanced Materials.

J.S and L.J.C contributed equally to this work.

\end{document}